\documentclass[sigconf]{acmart}
\usepackage[hyperref]{}

\usepackage{graphicx}
\graphicspath{ {images/} }

\usepackage[utf8]{inputenc}
\usepackage{longtable}
\usepackage{float} %to put figure on strict location - specific place [H]
\usepackage{booktabs} % For formal tables

\usepackage[english]{babel}
\usepackage{graphicx}
\usepackage{comment}
\usepackage{color}
\usepackage{xspace}
\usepackage{algorithm}
\usepackage{algpseudocode}
\usepackage{subcaption}
\usepackage{amsmath}
\usepackage{enumitem}
\usepackage[colorinlistoftodos]{todonotes}
\usepackage{mdwlist}

\begin{document}

% Copyright
%\setcopyright{acmcopyright}
%\setcopyright{acmlicensed}
%\setcopyright{rightsretained}
%\setcopyright{usgov}
%\setcopyright{usgovmixed}
%\setcopyright{cagov}
%\setcopyright{cagovmixed}

\setcopyright{acmcopyright} % if you give the rights to ACM
\acmDOI{...} % DOI - Insert your DOI below...
\acmISBN{...} % ISBN - Insert your conference/workshop's ISBN below...
\acmYear{2017} % Insert Publication year
\copyrightyear{2017} % Insert Copyright year (typically the same as above)
\acmPrice{15.00}
\acmConference[Short Name]{Long Name}{dates}{venue}
% DOI
%\doi{10.475/}

% ISBN
%\isbn{123-4567-24-567/08/06}

%Conference
%\conferenceinfo{}{}

%\acmPrice{\$15.00}

%
% --- Author Metadata here ---

%\CopyrightYear{2007} % Allows default copyright year (20XX) to be over-ridden - IF NEED BE.
%\crdata{0-12345-67-8/90/01}  % Allows default copyright data (0-89791-88-6/97/05) to be over-ridden - IF NEED BE.
% --- End of Author Metadata ---

%\title{Alternate {\ttlit ACM} SIG Proceedings Paper in LaTeX
\title{Assessing the Quality of Scientific Papers
%Predicting the impact of scientific articles 
%\titlenote{(Produces the permission block, and copyright information). For use with SIG-ALTERNATE.CLS. Supported by ACM.}
}
%\subtitle{[Extended Abstract]
%\titlenote{A full version of this paper is available as
%\textit{Author's Guide to Preparing ACM SIG Proceedings Using
%\LaTeX$2_\epsilon$\ and BibTeX} at

%  in this sample file, there are a *total*
\author{Roman Vainshtein}
\orcid{1234-5678-9012}
\affiliation{%
  \institution{Ben-Gurion University of the Negev}
  \streetaddress{P.O.Box 653}
  \city{Beer-Sheva}
  \state{Israel}
  \postcode{8410501}
}
\email{romanva@post.bgu.ac.il}

\author{Gilad Katz}
\affiliation{%
  \institution{Ben-Gurion University of the Negev}
  \streetaddress{P.O.Box 653}
  \city{Beer-Sheva}
  \state{Israel}
  \postcode{8410501}
}
\email{giladkz@post.bgu.ac.il}

\author{Bracha Shapira}
\affiliation{%
  \institution{Ben-Gurion University of the Negev}
  \streetaddress{P.O.Box 653}
  \city{Beer-Sheva}
  \state{Israel}
  \postcode{8410501}
}
\email{bshapira@post.bgu.ac.il}

\author{Lior Rokach}
\affiliation{%
  \institution{Ben-Gurion University of the Negev}
  \streetaddress{P.O.Box 653}
  \city{Beer-Sheva}
  \state{Israel}
  \postcode{8410501}
}
\email{liorrk@post.bgu.ac.il}

%#\renewcommand{\shortauthors}{B. Trovato et al.}

\begin{abstract}

A multitude of factors are responsible for the overall quality of scientific papers, including readability, linguistic quality, fluency, semantic complexity, and of course domain-specific technical factors. These factors vary from one field of study to another.
In this paper, we propose a measure and method for assessing the overall quality of the scientific papers in a particular field of study. We evaluate our method in the computer science domain, but it can be applied to other technical and scientific fields.

Our method is based on the corpus linguistics technique. This technique enables the extraction of required information and knowledge associated with a specific domain. For this purpose, we have created a large corpus, consisting of papers from very high impact conferences. First, we analyze this corpus in order to extract rich domain-specific terminology and knowledge. Then we use the acquired knowledge to estimate the quality of scientific papers by applying our proposed measure. We examine our measure on high and low scientific impact test corpora. Our results show a significant difference in the measure scores of the high and low impact test corpora. Second, we develop a classifier based on our proposed measure and compare it to the baseline classifier. Our results show that the classifier based on our measure over-performed the baseline classifier. Based on the presented results the proposed measure and the technique can be used for automated assessment of scientific papers.

%Firstly, we performed standard readability estimation on our corpora, in order to analyze any differences between high impact and low impact papers.
%Great writing is important factor on the way to achieve high impact in science research publications, but it's not the only factor.

\end{abstract}

\maketitle 

\begin{CCSXML}
	<ccs2012>
	<concept>
	<concept_id>10010405.10010497.10010504.10010505</concept_id>
	<concept_desc>Applied computing~Document analysis</concept_desc>
	<concept_significance>500</concept_significance>
	</concept>
	<concept>
	<concept_id>10002951.10003227.10003392</concept_id>
	<concept_desc>Information systems~Digital libraries and archives</concept_desc>
	<concept_significance>300</concept_significance>
	</concept>
	<concept>
	<concept_id>10002951.10003317.10003318.10003321</concept_id>
	<concept_desc>Information systems~Content analysis and feature selection</concept_desc>
	<concept_significance>300</concept_significance>
	</concept>
	
    <concept>
    <concept_id>10010147.10010178.10010179.10003352</concept_id>
    <concept_desc>Computing methodologies~Information extraction</concept_desc>
    <concept_significance>100</concept_significance>
    </concept>
    
	</ccs2012>
	
\end{CCSXML}

\ccsdesc[500]{Applied computing~Document analysis}
\ccsdesc[300]{Information systems~Digital libraries and archives}
\ccsdesc[300]{Information systems~Content analysis and feature selection}
\ccsdesc[300]{Computing methodologies~Information extraction}

\keywords{collocation; corpus linguistics; computational lexicography; lexical quality; log likelihood ratio; text; bag of words, classification}

\section{Introduction}
The meaning of the phrase "quality of writing" with regard to scientific papers is very broad, and it covers readability, linguistic quality, fluency, semantic complexity, clarity, comprehensibility, and domain-specific nuances such as the use of specialized technical terminology. Numerous studies have been conducted to explore text readability parameters and characteristics. The term "readability" is strongly linked with readability formulas which were introduced to measure the relative complexity and difficulty of texts.\cite{klare:read}Previous research has found that academic writing is classified by readability formulas into the "most difficult" levels\cite{gazni2011abstracts}. In other words, based on readability formulas it can be determined that, academic authors write in a way that is difficult to read.\cite{dolnicar2015readability}

However, readability formulas have limitations. The formulas tend to ignore readers' background knowledge, assumptions, interests, and motivations.\cite{bailin2016towards} Scientific and technical texts based on domain-specific knowledge, and they often consist of long and complex terms, making the texts incomprehensible to those outside the domain. In contrast, readers with prior domain knowledge, such as academic experts, are familiar with the terminology of their field and they will find scientific texts readable and understandable\cite{weir2007:optimising}. In addition, most of the classic readability formulas were developed from the 1940s to the 1970s, with an orientation toward manual use and a less technologically advanced society. In the computing era, attempts have been made to develop methods for automated evaluation of the linguistic quality and readability of texts. One of these methods is corpus linguistic, a growing discipline that applies analytical results from large language corpora to a wide variety of problems in linguistics and related domains.\cite{mcenery2011:corpus} A few directions have been developed that combining both readability formulas and the corpus linguistic technique\cite{anagnostou2006:corpus}. Other studies focusing on automated linguistic quality evaluation \cite{pitler:automatic} \cite{louis2013corpus} did not evaluate professional and domain-specific qualities of the text, which, in the case of technical documents, is a very important factor that determines the popularity and importance of the document in the particular technical field. Recent studies have attempted to predict the scientific impact of articles on the basis of their metadata\cite{dong2015predict_impact}. Such as the venue in which the paper is published, the researcher’s authority on the publication topic, co-authors, and more. To estimate the novelty and diversity of the research paper's topic, content analysis was done only based on the title and abstract. Other researches focusing on writing quality\cite{louis2013makes} achieving good results in text quality assessment when based only on content features. Although these researches show good potential in content analysis, they used very basic techniques to perform it.

\textbf{Contributions. }In this work, we focus on the automatic evaluation of the overall quality of scientific papers (linguistic and technical), and have developed a technique and measure for this purpose. Our measure provides an assessment of the overall quality of the text and articles on the basis of their content. We deal with linguistic quality, as well as the overall quality of the article in its domain. To date, no such measure has been introduced in other studies. The primary contribution of our work is a measure that can assess articles, even before they are published and without the use of metadata. We have created a measure that enables the automated filtering of high quality texts and articles in a specific domain, based only on examination of the papers’ content. 
We test our method on academic papers from the computer science domain. Our second contribution is a large dataset of collocations from the computer science domain, extracted from more that 30,000 high impact research papers. This dataset will be available for download by interested researchers for any potential future research.

%In natural language, words are not combined randomly. Although constrained by grammar and syntax words also have "preferences". Firth (1957) named such preferential word combinations "collocation".
%
%In modern scientific community it is used to evaluate the quality of research work based on the number of citations it has gained.
%
%The range of the vocabulary sampled from the text. the range was taken to mean the number of different words per thousand. 
%
%Psycholinguistic studies have shown that people read frequent words and phrases more quickly, so the words that appear in a text might influence people's perception of its quality.\cite{pitler:automatic} 
%Based on what is known we assume the "rule of thumb": If some article was cited a lot, then it's probably, clearly written and easily understandable article. 

\section{Related Work}
Despite the limitations of readability metrics, they are still used in the scientific community. They are easy to use and implement. Readability criteria is also the most widely accepted parameter used by the academic community to measure text complexity. Therefore, readability metrics may provide a baseline from which to extend our work with more intelligent metrics.
The most commonly used metrics in readability analysis are:
\begin{itemize}
  \item Flesch reading ease score \cite{flesch1949:Flesch}
  \item Flesh-Kincaid readability formula \cite{kincaid1975:FLESCH_KINCAID}
  \item Gunning fog index \cite{gunning1952:GUNNING_FOG}
  \item SMOG grading or index \cite{mc1969smog}
  \item Automated readability index \cite{senter1967:ARI}
  \item Coleman–Liau computer reliability formula \cite{coleman1975:Coleman_Liau}
\end{itemize}

While most of readability analysis was introduced before the computing era and can be calculated manually, there are also methods for automated evaluation of the linguistic quality and readability of texts like corpus linguistics.
Corpus linguistics is defined as the study of language based on examples of "real life" language use \cite{mcenery2001corpus}.
Corpus linguistics is not a branch of linguistics itself, rather it is a methodology and technique for language study which can be used in any branch of linguistics. 
In the area of corpus linguistics, a corpus is designed with a purpose in mind, which will define the corpus itself. In our work, the corpus serves to analyze academic papers; such a corpus is referred to as a "specialized corpus"(a corpus of texts of a specific type, for instance, scientific articles)\cite{anagnostou2006:corpus}, and these kinds of corpora are used to investigate a particular type of language \cite{chang2014:use_specialized}.
Such a corpus allow to perform linguistic studies in specific domain such as scientific computer science papers. 

The collocation extraction process is an important component of corpus linguistic studies. A collocation is an expression consisting of two or more words that correspond to a conventional way of saying things. Several studies have used the collocation frequency in the text as an indicator of semantic complexity\cite{anagnostou2007:average} \cite{gao2014:automatic} and as text readability measure\cite{anagnostou2006:corpus}.  

Our work is related to corpus linguistics, collocation extraction methods, and readability metrics. However, in contrast to previous studies, we do not use these methodologies for traditional linguistic tasks; instead, they are used, in order to develop a sophisticated measure for assessing the overall quality of technical papers.  

\section{Our Corpora}
As previously mentioned, in our work we are interested in assessing the overall quality of the text in computer science academic papers on the basis of corpus linguistics. For this purpose, the main corpus was created from selected articles from the most high impact IEEE\footnote{Institute of Electrical and Electronics Engineers} and ACM\footnote{Association for Computing Machinery} conferences. These conferences were selected on the basis of the CORE\footnote{The Computing Research and Education Association of Australasia, CORE} Conference Ranking which provides assessments of major conferences in the computing disciplines. The main corpus will serve as our "knowledge source" and as the "gold standard" of the quality of writing for computer science academic papers. Considering this fact, our selection criteria for papers was that they had to be published in conferences with A* or A rankings\footnote{CORE Conference Ranking -  http://www.core.edu.au/conference-portal}. The corpus consists of 60 conferences and almost 30,000 academic papers (see the complete corpus sources list in Appendix A).

In order to test our method, we created the test corpora.
Our test corpora are divided into two types: high impact and low impact conference papers. 

\begin{enumerate}  
	\item High Impact Papers Corpus
		\begin{enumerate}
			\item ACM Special Interest Group on Knowledge Discovery and Data Mining (SIGKDD) Conference - 490 papers
			\item ACM Special Interest Group on Information Retrieval (SIGIR) Conference - 406 papers
			\item IEEE International Conference on Data Mining (ICDM) - 1407 papers
		\end{enumerate}
	\item Low Impact Papers Corpus 
	\begin{enumerate}
		\item WSEAS Transactions on Computers Conference - 786 papers
		\item WSEAS European Computer Conference - 438 papers
		\item WSEAS International Conference on Artificial Intelligence, Knowledge Engineering and Data Bases - 494 papers
		\item EEEI Israel - 801 papers 
		\end{enumerate}
\end{enumerate}        
The impact ranking metrics that we used to select the conferences are based on the CORE Conference Rankings Portal and Google Scholar’s citation metrics (h5-median and h5-index)\footnote{https://scholar.google.com/intl/en/scholar/metrics.html}. Our test corpora papers and their impact rankings were gathered from publicly available web sources.
In summary, in this work we will deal with three corpora:
\begin{itemize}
	\item Main corpus - a large collection from high impact IEEE and ACM conferences.
	\item High impact papers (test) corpus - from SIGKDD, SIGIR and ICDM conferences.
	\item Low impact papers (test) corpus - from WSEAS and EEEI conferences. 
\end{itemize}

In Table 1, we present the impact scores of conferences from the test corpora (note that some of the low impact conferences were not ranked by Google Scholar or CORE, or were ranked by only one of them). 

\begin{table}[!h]
\centering
\caption{The impact scores of conferences from the test corpora}
\resizebox{0.81\textwidth}{!}{\begin{minipage}{\textwidth}
\label{my-label1}
\begin{tabular}{|l|c|c|c|}
\hline
\multicolumn{1}{|c|}{\textbf{Conference}} & \begin{tabular}[c]{@{}c@{}}\textbf{Google Scholar}\\ \textbf{h5-index}\end{tabular} & \begin{tabular}[c]{@{}c@{}}\textbf{Google Scholar}\\ \textbf{h5-median}\end{tabular} & \begin{tabular}[c]{@{}c@{}}\textbf{CORE}\\ \textbf{Ranking}\end{tabular} \\ \hline
ACM SIGKDD & 67 & 98 & A* \\ \hline
ICDM & 39 & 64 & A* \\ \hline
ACM SIGIR & 50 & 80 & A* \\ \hline
EEEI Israel & 11 & 13 & - \\ \hline
WSEAS AIKED & 6 & 8 & - \\ \hline
WSEAS ECC & 6 & 10 & C \\ \hline
WSEAS TCOMP & - & - & - \\ \hline
\end{tabular}
\end{minipage}}
\end{table}

\section{Collocation}
A collocation is a sequence of words or terms that co-occur more often than would be expected by chance. There is considerable overlap between the concept of collocation and notions such as terms, technical terms, and terminological phrases. As these might indicate, the latter three are commonly used when collocations are extracted from technical domains.\cite{manning1999:NLPcollocation}
In addition, it is important to understand that collocations can have different degrees of "strength".
"Collocation strength" is the measure of association between two terms, which evaluates whether the co-occurrence is coincidental or statistically significant. These association measure scores are used to rank the extracted collocations from the examined text.

Association measures are classified as follows:
\begin{itemize}
	\item Frequency-based measures (e.g., based on absolute and relative co-occurrence frequencies)
	\item Information-theoretic measures (e.g., mutual information, entropy)
	\item Statistical measures (e.g., chi-square, t-test, log-likelihood ratio, Sorensen–Dice coefficient)
\end{itemize}
The following are the most well known association measures used in collocation extraction\cite{anagnostou2006:corpus} \cite{manning1999:NLPcollocation}:
\begin{itemize}
\item T-Score
\item Mutual Information
\item Log-Likelihood Ratio
\item Dice Coefficient
\item Z-Score	
\end{itemize}
In our work, explore collocations as a means of compensating for the limitations of readability metrics in scientific and technical texts that require domain-specific knowledge. \cite{weir2007:optimising}. For our collocation extraction process, we applied the log-likelihood measure for the following reasons:
\begin{enumerate}
\item It is widely used and serves as the de facto common standard in computational linguistics for collocation extraction.
\item The majority of the NLP processing tools supports it.
\item It demonstrates good performance rates, and this is an important factor when dealing with large corpora.
\end{enumerate}

\section{Our Approach to Measuring the Quality of Scientific Texts}
Our method consists of several phases that the papers must pass through; the process is slightly different for the main corpus and the test corpora papers. The knowledge acquisition phase is carried out one time only, for the main corpus, and consists of:
\begin{enumerate}
    \item Preprocessing, which converts all scientific papers to plain text.
    \item Readability metrics extraction.
    \item Collocations and log-likelihood ratio score extraction and storage.
\end{enumerate}
The assessment phase for the test corpora consists of: 
\begin{enumerate}
    \item Preprocessing, which converts all scientific papers to plain text.
    \item Readability metrics extraction.
    \item Collocations extraction and storage.
    \item Calculation of the paper's quality score on the basis of our measure.
\end{enumerate}

Figure 1 demonstrates the entire process from the preprocessing phase to the final phase of assessing the test corpora. 
\begin{figure}[H]
\centering
\includegraphics[width=9cm]{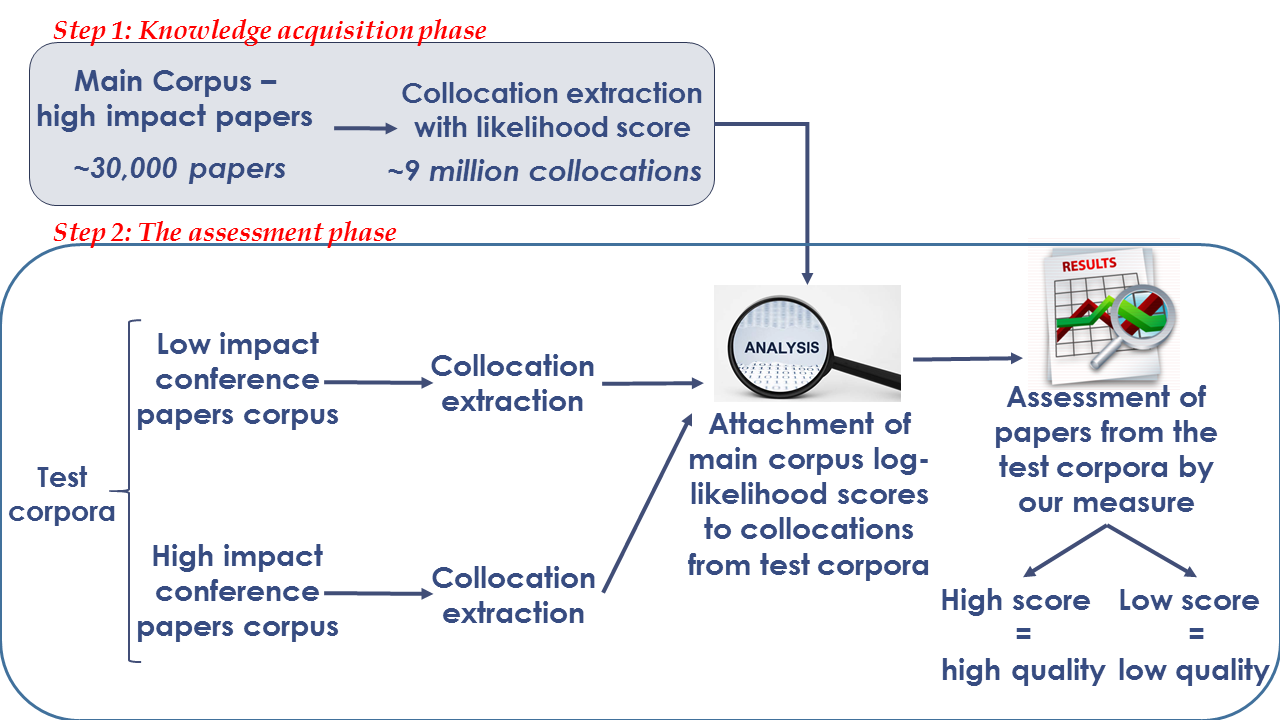}
\caption{The entire process of knowledge acquisition and test corpora assessment.}
\end{figure}

\subsection{Corpora Preprocessing}
As previously mentioned, we work with thee corpora: a main corpus, a high impact test corpus, and a low impact test corpus.
In the first phase, we performed preprocessing on each corpus. Preprocessing includes the conversion of each scientific paper from a PDF file to a plain text file.

\subsection{Readability Metric Extraction} 
In parallel with corpora preprocessing, we also analyzed each of the papers for readability metrics as: the Flesch reading ease score, the Flesh-Kincaid
readability formula, the Gunning fog index, the SMOG grading or index, the automated readability index, and the Coleman–Liau computer reliability formula. In addition, we gathered from each scientific paper in our corpora additional parameters, which summarize the number of: characters, syllables, words, complex words, sentences and commas, from each scientific paper in our corpora.

\subsection{Collocation Extraction}

For the collocation extraction phase, our team used the Natural Language Toolkit (NLTK)\footnote{http://www.nltk.org/}.
In this work, we focused only on bigram collocation extraction.
First, the main corpus was processed integrally, in order to remove all stop words and punctuation before collocation extraction took place. After extraction, all collocations, along with association measures, were stored in a database for use in the next phase.
In the second step, we extracted collocations from our test corpora, but this time we processed each paper individually and treated its collocations separately. Each collocation was stored in the database, along with the paper's DOI (digital object identifier or another ID string of the paper). 
As previously mentioned, our association measure to score collocations is the log likelihood ratio.
Tables 2 and 3 present examples of the main and test corpora collocations:

\begin{table}[!h]
\centering
\caption{Main corpus collocations extraction example}
\begin{tabular}{|c|c|c|} \hline
\textbf{Coll. word1}&\textbf{Coll. word2}&\textbf{Log-likelihood score}\\ \hline
source & code & 132755.20\\ \hline
augmented &  reality & 176330.41\\ \hline
training &  set & 92254.74\\ \hline
experimental &  results & 167023.14\\
\hline\end{tabular}
\end{table}

\begin{table}[!h]
\centering
\caption{Test corpus collocations extraction example}
 \resizebox{0.95\textwidth}{!}{\begin{minipage}{\textwidth}
\begin{tabular}{|c|c|c|} \hline
\textbf{DOI}&\textbf{Coll. word1}&\textbf{Coll. word2}\\ \hline
10.1109/ICDM.2009.19 & source & code\\ \hline
10.1109/ICDM.2009.19 & distributed & system \\ \hline
10.1109/EEEI.2006.321081 & state & vector \\ \hline
10.1109/EEEI.2006.321081 & artificial & data \\ \hline
\end{tabular}
\end{minipage}}
\end{table}

In total, we extracted more than nine million collocations from the main corpus (the full collocation list and scores are available for download from our website\footnote{https://goo.gl/1s4Jkx}).

\subsection{A Measure of the Paper's Quality}

Our main corpus is a large collection of papers from high impact conferences.
Collocations from our main corpus represent the common language in the computer science domain. The association measure score of each collocation (the log likelihood ratio) is a kind of index that expresses how widespread and accepted a collocation is in the computer science domain. In addition, this collocation association measure score at the same time can represents the rate of spelling errors in scientific paper.
In order to use this knowledge with our test corpora to measure a test paper's quality, we applied the log-likelihood scores from the main corpus collocations to any similar collocations that we found in our test paper. For example, as shown in Table 2, the collocation "source code" obtained a log-likelihood ratio score of 132,755.20 in the main corpus. Consequently, this score will be assigned to similar collocations in test paper that appear in Table 3. In our example, the "source code" collocation appears in the second row of Table 3.
In this method, we review the test paper collocations and assigned a main corpus log-likelihood ratio score to each of them.   
In the next step, we calculate the paper’s average log-likelihood ratio score:
\begin{equation}
\textit{ADS=}\frac{1}{m}\sum_{i=1}^{m} f_i
\end{equation}
In order to normalize the score based on the paper's length, we also added the following variation:

\begin{equation}\textit{ADSn=}\frac{1}{w}\sum_{i=1}^{m} f_i\end{equation}

Where:
\begin{itemize}
\item \textit{ADS} = Average document score
\item \textit{ADSn} = Average document score normalized based on the paper's length
\item \textit{m} = Number of different types of collocations in the considered paper that also appear in the main corpus
\item \textit{fi} = Log-likelihood score of collocation type i in the main corpus
\item \textit{w} = Total number of words in the considered paper
\end{itemize}

In this way, we can examine both of our test corpora, the high impact versus the low impact one. We assume that this measure will appropriately express the "quality" of papers from the main corpus domain that are not necessarily part of the computer science domain. Other scientific domains can be examined by changing the main corpus accordingly.

\section{Experimental Results}
First, we present our papers' quality measure experimental results for the high and low impact corpora.  

Second, we present experimental results for the readability metrics, which examine our main and test corpora from a complexity perspective.

\subsection{The Papers' Quality Measure Results}
In order to verify and prove our measure, we examined it on our test corpora. Our hypothesis is that high impact corpus papers will obtain a higher score, and consequently, low impact corpus papers will obtain lower scores.
All 2,320 papers from high impact conferences (SIGKDD, SIGIR, ICDM) and 2,519 papers from low impact conferences (WSEAS-ECC, WSEAS TCC, WSEAS AIKED and EEEI Israel) were accessed with our measure (see Table 4).

A one-way between subjects ANOVA was conducted to compare the results for each of the conferences. There was a significant difference between the results at the p<.05 level for all seven conferences tested [F(6, 4874) = 29.59091502, p = 5.09153E-35] for both the non-normalized measure version and [F(6, 4874) = 6.960458274, p = 2.20342E-07] for the normalized version of the measure.
\begin{table}[!h]
\centering
\caption{Summary of the results with the normalized measure version}

\label{my-label7}
 \resizebox{0.80\textwidth}{!}{\begin{minipage}{\textwidth}
\begin{tabular}{|l|c|c|c|c|}
\hline
\multicolumn{1}{|c|}{\textbf{Groups}} & \textbf{Papers Count} & \textbf{Sum} & \textbf{Average} & \textbf{Variance} \\ \hline
EEEI Israel & 801 & 339966 & 424 & 95813 \\ \hline
WSEAS AIKED & 494 & 241274 & 488 & 75373 \\ \hline
WSEAS ECC & 438 & 219494 & 457 & 73119 \\ \hline
WSEAS TCOMP & 786 & 371751 & 473 & 40624 \\ \hline
SIGKDD & 490 & 317184 & \textbf{647}   & 189757 \\ \hline
SIGIR & 406 & 263153 & \textbf{648}  & 109463 \\ \hline
ICDM & 1407 & 1162921 & \textbf{827} & 105657 \\ \hline
\end{tabular}
\end{minipage}}
\end{table}
In order to reduce post hoc comparisons, we treated the high and low impact conferences papers as two distinct groups.
This enabled us to conduct an independent samples t-test to compare the high and low impact groups.
As can be observed, there was a significant difference between the high impact papers' measure scores (Mean=5949, SD=8396) and the low impact papers' measure scores (mean=3991, SD=2684); t(3538) = 10.75, p = 2.00231E-26, for the non-normalized measure. For the normalized measure version, the results, in principle, were analogous to the non-normalized version. In this case, we also observed a significant difference between the high impact papers’ measure scores (mean=898, SD=4066) and the low impact papers' measure scores (mean=458, SD=267); t(2337) = 5.21, p = 2.08575E-07 (see Table 5).

\begin{table}[!h]
\centering
\caption{Independent samples t-test for high and low impact test corpora - normalized measure version}
\label{my-label9}
\begin{tabular}{|l|c|c|}
\hline
 & \multicolumn{1}{l|}{\textbf{High Impact}} & \multicolumn{1}{l|}{\textbf{Low Impact}} \\ \hline
Mean & 757 & 458 \\ \hline
Variance & 131703 & 71195 \\ \hline
Observations & 2303 & 2519 \\ \hline
Hypothesized Mean Diff. & \multicolumn{2}{c|}{0} \\ \hline
df & \multicolumn{2}{c|}{4193} \\ \hline
t Stat & \multicolumn{2}{c|}{32.45} \\ \hline
P(T\textless=t) one-tail & \multicolumn{2}{c|}{1.4955E-206} \\ \hline
t Critical one-tail & \multicolumn{2}{c|}{1.65} \\ \hline
P(T\textless=t) two-tail & \multicolumn{2}{c|}{2.991E-206} \\ \hline
t Critical two-tail & \multicolumn{2}{c|}{1.96} \\ \hline
\end{tabular}
\end{table}

Our results suggest that the proposed measure can identify, and therefore "reward" (with a significantly higher score), the papers from high impact conferences, compared to papers from low impact conferences, which obtain lower scores.

It is important to note the role of the main corpus in our method. As follows from our preliminary experiments, if the main corpus is not built on the basis of very high impact conferences papers, the results obtained are not significant. Thus, the measure will not be effective in this case.

\subsection{Readability Metric Results}

In previous sections, we explained the importance of readability metrics and introduced the metrics that are analyzed in our work; the readability metric results for the main and test corpora appear in Table 6.


\begin{table}[!h]
\centering
\caption{Readability metric results - main and test corpora}
\label{my-label6}
 \resizebox{0.70\textwidth}{!}{\begin{minipage}{\textwidth}
\begin{tabular}{|l|c|c|c|l|}
\hline

\multicolumn{1}{|c|}{\textbf{\begin{tabular}[c]{@{}c@{}}Readability \\ Metric\end{tabular}}} & \textbf{\begin{tabular}[c]{@{}c@{}}Main\\ Corpus\end{tabular}} & \textbf{\begin{tabular}[c]{@{}c@{}}High Impact\\ Test Corpora\end{tabular}} & \textbf{\begin{tabular}[c]{@{}c@{}}Low Impact\\ Test Corpora\end{tabular}} & \multicolumn{1}{c|}{\textbf{Comment}} \\ \hline
SMOG & 14.46 & 14.36 & 15.71 & Less is easier \\ \hline
\begin{tabular}[c]{@{}l@{}}FLESCH\\  READING\end{tabular} & 40.12 & 42.82 & 33.34 & More is easier \\ \hline
\begin{tabular}[c]{@{}l@{}}FLESCH\\  KINCAID\end{tabular} & 13.15 & 12.55 & 14.93 & Less is easier \\ \hline
ARI & 12.47 & 11.61 & 14.57 & Less is easier \\ \hline
\begin{tabular}[c]{@{}l@{}}GUNNING\\  FOG\end{tabular} & 16.29 & 15.60 & 18.24 & Less is easier \\ \hline
\begin{tabular}[c]{@{}l@{}}COLEMAN \\ LIAU\end{tabular} & 11.72 & 11.20 & 12.35 & Less is easier \\ \hline
SMOG INDEX & 13.87 & 13.77 & 15.07 & Less is easier \\ \hline
CHARACTERS & 25588.91 & 27990.90 & 18621.00 &  \\ \hline
SYLLABLES & 9014.58 & 9888.51 & 6556.58 &  \\ \hline
WORDS & 5349.77 & 5898.98 & 3753.81 &  \\ \hline
\begin{tabular}[c]{@{}l@{}}COMPLEX \\ WORDS\end{tabular} & 1012.90 & 1093.94 & 793.94 &  \\ \hline
SENTENCES & 271.30 & 290.35 & 171.00 &  \\ \hline
COMMAS & 352.81 & 414.31 & 287.67 &  \\ \hline
\end{tabular}%
\end{minipage}}
\end{table}
 
An independent samples t-test was conducted to compare SMOG readability metrics scores of high and low impact test corpora (other metrics are very similar and are based on the same or similar parameters; for this reason we did not perform t-tests for other metrics). There was a significant difference between the low impact scores (mean=15.71, SD=4.02) and the high impact scores (mean=14.36, SD=1.79); t(3538) = 15.32, p = 2.45E-51. These results suggest that the high impact papers are more readable than the low impact ones.
This proves to be a very unusual result, differing from the results of previous researches. \cite{gazni2011abstracts}\cite{dolnicar2015readability}
Another interesting observation based on our research is that high impact papers are much longer and include more words than low impact articles.
We must, however, note that readability metrics results only indicate the complexity of papers, and not their quality, and therein is the limitation of these metrics.
It is important to note, that when we examined specific high and low impact conferences the difference wasn't significant in all cases (See table 8). Another point to pay the attention is the high values of the variance in low impact papers.
Based on these results, we can conclude, that readability metrics can not serve as a good measure of the paper's quality.
However, we can assume that readability metrics show that, less experienced authors are not able to express their thoughts in a detailed manner and in simple language.

\begin{table}[!h]
\centering
\caption{Independent-samples t-test - SMOG (test corpora)}
\resizebox{0.95\textwidth}{!}{\begin{minipage}{\textwidth}
\label{my-label12}
\begin{tabular}{|l|c|c|}
\hline
 & \textbf{Low Impact} & \textbf{High Impact} \\ \hline
Mean & 15.71 & 14.36 \\ \hline
Variance & 16.18 & 3.20 \\ \hline
Std. Deviation & 4.02 & 1.79  \\ \hline
Observations & 2519 & 2303 \\ \hline
Hypothesized Mean Diff. & \multicolumn{2}{c|}{0} \\ \hline
df & \multicolumn{2}{c|}{3538} \\ \hline
t Stat & \multicolumn{2}{c|}{15.32} \\ \hline
P(T\textless=t) one-tail & \multicolumn{2}{c|}{1.23E-51} \\ \hline
t Critical one-tail & \multicolumn{2}{c|}{1.65} \\ \hline
P(T\textless=t) two-tail & \multicolumn{2}{c|}{2.45E-51} \\ \hline
t Critical two-tail & \multicolumn{2}{c|}{1.96} \\ \hline
\end{tabular}
\end{minipage}}
\end{table}

\begin{table}[!h]
\centering
\caption{Independent-samples t-test - SMOG (ICDM and EEEI Israel papers)}
\resizebox{0.95\textwidth}{!}{\begin{minipage}{\textwidth}
\label{my-label5}
\begin{tabular}{|l|c|c|}
\hline
 & \textbf{\begin{tabular}[c]{@{}c@{}}EEEI\\ Low Impact\end{tabular}} & \textbf{\begin{tabular}[c]{@{}c@{}}ICDM\\ High Impact\end{tabular}} \\ \hline
Mean & 14.90 & 14.40 \\ \hline
Variance & 21.40 & 4.35 \\ \hline
Std. Deviation & 4.63 & 2.09  \\ \hline
Observations & 801 & 1407 \\ \hline
Hypothesized Mean Diff. & \multicolumn{2}{c|}{0} \\ \hline
df & \multicolumn{2}{c|}{986} \\ \hline
t Stat & \multicolumn{2}{c|}{-2.88} \\ \hline
P(T\textless=t) one-tail & \multicolumn{2}{c|}{0.002} \\ \hline
t Critical one-tail & \multicolumn{2}{c|}{1.646} \\ \hline
P(T\textless=t) two-tail & \multicolumn{2}{c|}{0.004} \\ \hline
t Critical two-tail & \multicolumn{2}{c|}{1.962} \\ \hline
\end{tabular}
\end{minipage}}
\end{table}

\subsection{Statistical Classification Based On Quality Measure And Readability Metrics}
To evaluate the performance of our metric in practice and subsequently compare it with other techniques, we developed a machine learning classifier based on our measure (Section 5.4).
First, we added to the test corpora from Table 1 a class feature (target feature), a 'High' or 'Low' value according to the type of conference the paper appeared in. As we mentioned, our test corpora consisted of 4822 papers in total (2303 high impact and 2519 low impact papers).
Second, our test corpora was divided to the train and test set. We used the Python SciKit-Learn Library to split the test corpora into random train and test subsets with train size of 80\% and test size of 20\% of the corpora. 
Thirdly, we used the Logistic Regression (aka logit, MaxEnt) classifier in Python SciKit-Learn Library to learn and predict the target feature for each paper ('High' or 'Low' class). 
In order to evaluate our trained classifier, we computed a receiver operating characteristic (ROC) and area under the curve (AUC) from prediction scores.
We evaluate four versions of the classifier. In each version, we changed or enlarged the training vector with more metrics (features).
As a basic try, we used only readability metrics (the first seven rows in Table 6) as the training vector for each paper (as mentioned before the target vector was the impact - 'High' or 'Low'). This classifier achieved an AUC score of 0.67 (AUC=0.67).
Our second try was to "enrich" our training vector with basic statistical properties of the paper, such as number of characters in the paper, syllables, words, complex words, sentences and commas (the last six rows in Table 6).
This classifier achieved an AUC score of 0.77 (AUC=0.77).
In our third classifier, we used only the proposed papers' quality measure scores (paragraph 5.4); this classifier achieved an AUC score of 0.80 (AUC=0.80).
In our final version of the classifier we unified the readability, the statistical properties and the papers' quality measure score as the training vector.
The final version achieved an AUC score of 0.85 (AUC=0.85).

\subsection{Baseline Comparisons With Bag Of Words Model}
In the previous section we examined the performance of a classifier that was based on the paper’s quality measure, readability and statistical metrics.
To compare the performance of our classifiers to other basic classification techniques, we used the Bag of Words model classifier technique as a baseline. This technique has proved to be very effective and simple for article quality
prediction in the science journalism domain, as presented in Louis and Nenkova work (Louis and Nenkova, 2013).
Due to a limitation on the quantity of low impact conference documents, we used only our test corpora for this task, both for training and for testing.
In order to simulate the same conditions as for previous classifiers that were trained on a variety of conferences with different topics, our train and test sets did not include papers from identical conferences.
For our train set we randomly chose a sample of 3,000 papers from our main high impact corpus (almost 30,000 docs) and 1,718 papers from the low impact corpora (WSEAS AIKED - 494, WSEAS TCOMP - 786, WSEAS ECC - 438). See Table 9.
Our test set contained 801 papers from EEEI Israel and a randomly chosen sample of 801 papers from the ICDM, KDD and SIGIR conferences. See Table 10.
For our bag of words model classifier training process, we used a tf-idf (term-frequency times inverse document-frequency) representation of the document.
The goal of using tf-idf is to scale down the impact of tokens that occur very frequently in a given corpus, and that are hence empirically less informative than features that occur in a small fraction of the train set.
For the classification task, we worked with a linear Support Vector Machine (SVM) algorithm on the basis of SciKit-Learn Library SGDClassifier implementation.

In the same way as in the previous section, in order to evaluate our trained classifier we used a Compute Receiver operating characteristic (ROC) and Area Under the Curve (AUC) from prediction scores.
The bag of words SVM classifier achieved an AUC score of 0.57 (AUC=0.57).

These results may indicate that the topic of the article plays an important role in article classification with the bag of words model. In other words, it is possible that, due to the difference in each of the conference's topics between EEEI and WSEAS, we got these results.

In order to verify whether the previous statement has some basis, we decided to change our train and test sets.
Our low impact train set was changed to: EEEI Israel - 801 papers, WSEAS AIKED - 494 papers and WSEAS ECC - 438 papers, with a total of 1733 documents. The high impact train set wasn't changed.
The low impact test set was also changed to WSEAS TCOMP - 786 papers. The high impact test set wasn't changed.
See Table 11 and 12.
In this configuration, the low impact test set is much closer in terms of topics to the train set, and vice versa.
This time our bag of words SVM classifier achieved an AUC score of 0.74 (AUC=0.74).
These results reinforce our statement about the important role of topics in article classification with the bag of words model.
In addition, it seems that the classifier based on our measure has an advantage, due to the fact that it does not depend so strongly on the topic of documents, but rather in the domain in a very general manner. The classifier based on our measure over-performed the baseline classifier, and in combination with readability and statistical metrics, achieved the best results. 
Table 13 presents a summary of the AUC results of all presented classifiers.

\begin{table}[!h]
\centering
\caption{Bag of words model classifier's train set - version 1}
\resizebox{0.95\textwidth}{!}{\begin{minipage}{\textwidth}
\label{my-label10}
\begin{tabular}{|l|c|c|}
\hline
\multicolumn{3}{|c|}{\textbf{Train Set}}                                          \\ \hline
\textbf{Conferences} & \begin{tabular}[c]{@{}c@{}}WSEAS AIKED \\ + WSEAS TCOMP \\ + WSEAS ECC\end{tabular} & High Impact corpus \\ \hline
\textbf{Class/Impact} & Low                                  & High               \\ \hline
\textbf{Documents}    & 1718                                 & 3000               \\ \hline
\textbf{Total}        & \multicolumn{2}{c|}{4718}                                 \\ \hline
\end{tabular}
\end{minipage}}
\end{table}

\begin{table}[]
\centering
\caption{Bag of words model classifier's test set - version 1}
\resizebox{1\textwidth}{!}{\begin{minipage}{\textwidth}
\label{my-label4}
\begin{tabular}{|l|c|c|}
\hline
\multicolumn{3}{|c|}{\textbf{Test Set}}                  \\ \hline
\textbf{Conferences}  & EEEI Israel & ICDM + KDD + SIGIR \\ \hline
\textbf{Class/Impact} & Low         & High               \\ \hline
\textbf{Documents}    & 801         & 801                \\ \hline
\textbf{Total}        & \multicolumn{2}{c|}{1602}        \\ \hline
\end{tabular}
\end{minipage}}
\end{table}

\begin{table}[]
\centering
\caption{Bag of words model classifier's train set - version 2}
\resizebox{0.92\textwidth}{!}{\begin{minipage}{\textwidth}
\label{my-label3}
\begin{tabular}{|l|c|c|}
\hline
\multicolumn{3}{|c|}{\textbf{Train Set}}                                                                                         \\ \hline
\textbf{Conferences}  & \begin{tabular}[c]{@{}c@{}}WSEAS AIKED \\ + WSEAS ECC \\ + EEEI Israel\end{tabular} & ICDM + KDD + SIGIR \\ \hline
\textbf{Class/Impact} & Low                                                                                 & High               \\ \hline
\textbf{Documents}    & 1733                                                                                & 3000               \\ \hline
\textbf{Total}        & \multicolumn{2}{c|}{4733}                                                                                \\ \hline
\end{tabular}
\end{minipage}}
\end{table}

\begin{table}[]
\centering
\caption{Bag of words model classifier's test set - version 2}
\resizebox{0.9\textwidth}{!}{\begin{minipage}{\textwidth}
\label{my-label}
\begin{tabular}{|l|c|c|}
\hline
\multicolumn{3}{|c|}{\textbf{Test Set}}                  \\ \hline
\textbf{Conferences}  & WSEAS TCOMP & ICDM + KDD + SIGIR \\ \hline
\textbf{Class/Impact} & Low         & High               \\ \hline
\textbf{Documents}    & 786         & 801                \\ \hline
\textbf{Total}        & \multicolumn{2}{c|}{1587}        \\ \hline
\end{tabular}
\end{minipage}}
\end{table}

\begin{table}[!htbp]
\centering
\caption{Summary of all classifiers - AUC results}
\resizebox{0.95\textwidth}{!}{\begin{minipage}{\textwidth}
\label{my-label2}
\begin{tabular}{|l|c|}
\hline
\multicolumn{1}{|c|}{\textbf{Classifier based on:}} & \textbf{AUC Result} \\ \hline
Readability metrics                                 & 0.67                \\ \hline
Readability metrics + basic doc. statistic.         & 0.77                \\ \hline
Purposed quality measure                            & 0.80                \\ \hline
Readability + statistic. + quality measure          & 0.85                \\ \hline
Bag of words (baseline) - ver 1                     & 0.57                \\ \hline
Bag of words (baseline) - ver 2                     & 0.74                \\ \hline
\end{tabular}
\end{minipage}}
\end{table}

\section{Conclusions and Future Work}
Based on our findings and results, the proposed measure can be used to predict the "quality" and impact of a scientific paper.  In order to expand this work, it would be very helpful to collaborate with high impact conference organizers, who can provide additional test corpora consisting of the rejected papers and those that were accepted for the conference.
In this research, we constructed our main corpus, which served as our "knowledge source" from nearly 30,000 scientific papers from the most highly rated conferences in the computer science domain. Despite such a large number of papers, this number was limited by the NLTK tool and other computer resources. It would be interesting to explore a larger corpus created from an even larger number of papers. Another issue to explore is the domain variance of the main corpus; for instance, instead of the computer science domain, the corpus could be changed to another domain. The main corpus can also be restricted to a very specific field or widened to include more than one field. 
Another parameter that can affect the research is the collocation type. In our research, we extracted only bigram collocations, but the addition of trigram and fourgram collocations may improve the results and enrich the "knowledge".  
If the primary goal is to estimate the influence of the paper in a particular technical field, we also need to exclude common "knowledge" or common language collocations that likely affect the results. Creating a list of common collocations (from non-technical corpora) and excluding this group from the extracted collocations may improve the method, and accordingly the results. An additional direction that can be examined in the future is stemming, an area which was not performed in this research. 

%\end{document}  % This is where a 'short' article might terminate

\bibliographystyle{abbrv}
\bibliography{sigproc}
%\nocite{*}  % sigproc.bib is the name of the Bibliography in this case
% ACM needs 'a single self-contained file'!
%\balancecolumns
\appendix
%Appendix A

\section{Main corpus sources(the corpus serves as the "gold standard")}

\begin{table*}
\tiny
	\centering
	\caption{Main corpus sources, the corpus serves as "Gold standard"}
	\label{my-label11}
	\resizebox{10cm}{12cm} {
	\begin{tabular}{|p{4cm} |p{2cm} |p{1cm}|p{1.5cm}|p{3cm}|p{1cm}|}
		\hline
		\textbf{Conference}                                                                                                                                 & \textbf{Acronym}          & \textbf{Rank} & \textbf{Rank Source:} & \textbf{Field Of Research}                                                                   & \textbf{Papers Quantity} \\ \hline
		IEEE Symposium on Foundations of Computer Science                                                                                                   & FOCS                      & A*            & CORE2014              & 0802 - Computation Theory and Mathematics                                                    & 809                      \\ \hline
		IEEE International Conference on Computer Vision                                                                                                    & ICCV                      & A*            & CORE2014              & 0801 - Artificial Intelligence and Image Processing                                          & 2596                     \\ \hline
		IEEE Symposium on Logic in Computer Science                                                                                                         & LICS                      & A*            & CORE2014              & 0802 - Computation Theory and Mathematics                                                    & 1148                     \\ \hline
		IEEE International Conference on Pervasive Computing and Communications                                                                             & PERCOM                    & A*            & CORE2014              & 0805 - Distributed Computing                                                                 & 603                      \\ \hline
		IEEE Symposium on Security and Privacy                                                                                                              & SP (S\&P)                 & A*            & CORE2014              & 0802 - Computation Theory and Mathematics0803 - Computer Software                            & 263                      \\ \hline
		IEEE Information Visualization Conference                                                                                                           & INFVIS,VIS (IEEE InfoVis) & A*            & CORE2014              & 0801 - Artificial Intelligence and Image Processing0806 - Information Systems                & 316                      \\ \hline
		IEEE/ACM International Symposium on Mixed and Augmented Reality                                                                                     & ISMAR,ISMAR-AMH(ISMAR)    & A*            & CORE2014              & 0801 - Artificial Intelligence and Image Processing                                          & 856                      \\ \hline
		IEEE International Symposium on Wearable Computing                                                                                                  & ISWC                      & A*            & CORE2014              & 0806 - Information Systems                                                                   & 526                      \\ \hline
		IEEE Computer Security Foundations Symposium (was CSFW)                                                                                             & CSF, CSFW                 & A             & CORE2014              & 0803 - Computer Software                                                                     & 319                      \\ \hline
		IEEE International Conference on Web Services                                                                                                       & ICWS                      & A             & CORE2014              & 0806 - Information Systems                                                                   & 1312                     \\ \hline
		IEEE Conference on Computer Vision and Pattern Recognition                                                                                          & CVPR, CVPRW (CVPR)        & A             & CORE2014              & 0801 - Artificial Intelligence and Image Processing                                          & 6004                     \\ \hline
		IEEE International Conference on Services Computing                                                                                                 & SCC                       & A             & CORE2014              & 0806 - Information Systems                                                                   & 1109                     \\ \hline
		IEEE International Conference on Engineering of Complex Computer Systems                                                                            & ICECCS                    & A             & CORE2014              & 1005 - Communications Technologies                                                           & 685                      \\ \hline
		IEEE/IFIP Working Conference on Software Architecture                                                                                               & WICSA                     & A             & CORE2014              & 0803 - Computer Software                                                                     & 345                      \\ \hline
		IEEE Symposium on Computational Complexity                                                                                                          & CCC                       & A             & CORE2014              & 0802 - Computation Theory and Mathematics                                                    & 573                      \\ \hline
		IEEE Workshop on Applications of Computer Vision                                                                                                    & WACV                      & A             & CORE2014              & 0801 - Artificial Intelligence and Image Processing                                          & 616                      \\ \hline
		IEEE International Conference on Software Maintenance and Evolution & ICSME,ICSM (ICSME)        & A             & CORE2014              & 0803 - Computer Software                                                                     & 1485                     \\ \hline
		IEEE International Conference on Document Analysis and Recognition                                                                                  & ICDAR                     & A             & CORE2014              & 0806 - Information Systems                                                                   & 2362                     \\ \hline
		IEEE International Conference on Computer Languages                                                                                                 & ICCL                      & A             & CORE2014              & 0803 - Computer Software                                                                     & 172                      \\ \hline
		IEEE International Symposium on Software Metrics                                                                                                    & SMS                       & A             & CORE2014              & 0803 - Computer Software                                                                     & 46                       \\ \hline
		IEEE International Conference on Multimedia Computing and Systems                                                                                   & ICMCS                     & A             & CORE2014              & 0806 - Information Systems                                                                   & 730                      \\ \hline
		IEEE International Symposium on Cluster, Cloud and Grid Computing                                                                                   & CCGRID                    & A             & CORE2014              & 0805 - Distributed Computing                                                                 & 1395                     \\ \hline
		ACM Conference on Computer and Communications Security                                                                                              & CCS                       & A*            & CORE2014              & 0803 - Computer Software                                                                     & 205                      \\ \hline
		ACM International Symposium on Computer Architecture                                                                                                & ISCA                      & A*            & CORE2014              & 0803 - Computer Software                                                                     & 35                       \\ \hline
		ACM Conference on Digital Libraries                                                                                                                 & JCDL                      & A*            & CORE2014              & 0806 - Information Systems                                                                   & 63                       \\ \hline
		ACM Conference on Object Oriented Programming Systems Languages and Applications                                                                    & OOPSLA                    & A*            & CORE2014              & 0803 - Computer Software                                                                     & 69                       \\ \hline
		ACM-SIGPLAN Conference on Programming Language Design and Implementation                                                                            & PLDI                      & A*            & CORE2014              & 0803 - Computer Software                                                                     & 52                       \\ \hline
		ACM Symposium on Principles of Distributed Computing                                                                                                & PODC                      & A*            & CORE2014              & 0805 - Distributed Computing                                                                 & 125                      \\ \hline
		ACM SIGMOD-SIGACT-SIGART Conference on Principles of Database Systems                                                                               & PODS                      & A*            & CORE2014              & 0804 - Data Format                                                                           & 45                       \\ \hline
		ACM-SIGACT Symposium on Principles of Programming Languages                                                                                         & POPL                      & A*            & CORE2014              & 0803 - Computer Software                                                                     & 69                       \\ \hline
		ACM Conference on Applications, Technologies, Architectures, and Protocols for Computer Communication                                               & SIGCOMM                   & A*            & CORE2014              & 0803 - Computer Software                                                                     & 94                       \\ \hline
		ACM SIG International Conference on Computer Graphics and Interactive Techniques                                                                    & SIGGRAPH                  & A*            & CORE2014              & 0801 - Artificial Intelligence and Image Processing                                          & 405                      \\ \hline
		ACM SIG on Computer and Communications Metrics and Performance                                                                                      & SIGMETRICS                & A*            & CORE2014              & 1006 - Computer Hardware                                                                     & 45                       \\ \hline
		ACM Special Interest Group on Management of Data Conference                                                                                         & SIGMOD                    & A*            & CORE2014              & 0804 - Data Format                                                                           & 309                      \\ \hline
		ACM/SIAM Symposium on Discrete Algorithms                                                                                                           & SODA                      & A*            & CORE2014              & 0802 - Computation Theory and Mathematics                                                    & 1                        \\ \hline
		ACM SIGOPS Symposium on Operating Systems Principles                                                                                                & SOSP                      & A*            & CORE2014              & 0803 - Computer Software                                                                     & 48                       \\ \hline
		ACM Symposium on Theory of Computing                                                                                                                & STOC                      & A*            & CORE2014              & 0802 - Computation Theory and Mathematics                                                    & 304                      \\ \hline
		ACM International Conference on Web Search and Data Mining                                                                                          & WSDM                      & A*            & CORE2014              & 0804 - Data Format                                                                           & 168                      \\ \hline
		ACM Conference on Computer Supported Cooperative Work                                                                                               & CSCW                      & A             & CORE2014              & 0806 - Information Systems                                                                   & 122                      \\ \hline
		ACM Multimedia                                                                                                                                      & ACMMM (MM) (Multimedia)   & A*            & CORE2014              & 0803 - Computer Software1203 - Design Practice and Management                                & 308                      \\ \hline
		ACM International Conference on Information and Knowledge Management                                                                                & CIKM                      & A             & CORE2014              & 0806 - Information Systems 0807 - Library and Information Studies                            & 432                      \\ \hline
		ACM International Conference on Supercomputing                                                                                                      & ICS                       & A             & CORE2014              & 0805 - Distributed Computing                                                                 & 63                       \\ \hline
		ACM Special Interest Group on Computer Science Education Conference                                                                                 & SIGCSE                    & A             & CORE2014              & 0899 - Other Information and Computing Sciences                                              & 94                       \\ \hline
		ACM Symposium on Computational Geometry                                                                                                             & SCG                       & A             & CORE2014              & 0802 - Computation Theory and Mathematics                                                    & 65                       \\ \hline
		ACM Symposium on User Interface Software and Technology                                                                                             & UIST                      & A             & CORE2014              & 0806 - Information Systems                                                                   & 85                       \\ \hline
		ACM/IEEE Supercomputing Conference                                                                                                                  & SC                        & A             & CORE2014              & 0805 - Distributed Computing                                                                 & 50                       \\ \hline
		ACM Conference on Embedded Software                                                                                                                 & EMSOFT                    & A             & CORE2014              & 1005 - Communications Technologies                                                           & 20                       \\ \hline
		ACM Conference on Hypertext and Hypermedia                                                                                                          & Hypertext                 & A             & CORE2014              & 0805 - Distributed Computing0806 - Information Systems                                       & 26                       \\ \hline
		ACM International Symposium on High Performance Distributed Computing                                                                               & HPDC                      & A             & CORE2014              & 0805 - Distributed Computing                                                                 & 29                       \\ \hline
		Internet Measurement Conference                                                                                                                     & IMC                       & A             & CORE2010              & 1005 - Communications Technologies                                                           & 63                       \\ \hline
		ACM Virtual Reality Software and Technology                                                                                                         & VRST                      & A             & CORE2014              & 0801 - Artificial Intelligence and Image Processing                                          & 17                       \\ \hline
		IEEE/ACM International Conference on Computer-Aided Design                                                                                          & ICCAD                     & A             & CORE2014              & 0801 - Artificial Intelligence and Image Processing                                          & 40                       \\ \hline
		ACM International Conference on Human Factors in Computing Systems                                                                                  & CHI                       & A*            & CORE2014              & 0806 - Information Systems1203 - Design Practice and Management                              & 637                      \\ \hline
		ACM Genetic and Evolutionary Computations                                                                                                           & GECCO                     & A             & CORE2014              & 08 - Information and Computing Sciences09 - Engineering                                      & 343                      \\ \hline
		ACM International Conference on Human Factors in Computing Systems Extended Abstracts                                                               & CHI EA                    & A*            & CORE2014              & 0806 - Information Systems1203 - Design Practice and Management                              & 196                      \\ \hline
		ACM Design Automation Conf                                                                                                                          & DAC                       & A             & CORE2014              &                                                                                              & 364                      \\ \hline
		ACM International World Wide Web Conference                                                                                                         & WWW                       & A*            & CORE2014              & 0805 - Distributed Computing0806 - Information Systems0807 - Library and Information Studies & 260                      \\ \hline
		ACM International Conference on Software Engineering                                                                                                & ICSE                      & A*            & CORE2014              & 0803 - Computer Software                                                                     & 152                      \\ \hline
		ACM International Conference on Intelligent User Interfaces                                                                                         & IUI                       & A*            & CORE2014              & 0806 - Information Systems                                                                   & 69                       \\ \hline
		ACM International Conference on Machine Learning                                                                                                    & ICML                      & A*            & CORE2014              & 0801 - Artificial Intelligence and Image Processing                                          & 106                      \\ \hline
		\textbf{Total Papers (in corpus)}                                                                                                                   & \textbf{}                 & \textbf{}     & \textbf{}             & \textbf{}                                                                                    & \textbf{29848}           \\ \hline
	\end{tabular}
}
\end{table*}

%\subsection{References}
%Generated by bibtex from your ~.bib file.  Run latex,
%then bibtex, then latex twice (to resolve references)
%to create the ~.bbl file.  Insert that ~.bbl file into
%the .tex source file and comment out
%the command \texttt{{\char'134}thebibliography}.

\end{document}